\documentclass[11pt,twoside]{article}
\usepackage{asp2010}

\resetcounters

\begin{document}

\title{The Distribution of Warm Ionized Medium in Galaxies}
\author{L.M.~Haffner
\affil{Department of Astronomy, University of Wisconsin---Madison,\\475 North Charter Street, Madison, WI 53706}}

\begin{abstract}
Ionized nebulae have been targets of interest since the introduction of the telescope centuries ago. These isolated, ``classical'' H~\textsc{ii} regions gave us some of the earliest insight into the copious feedback energy that stars inject into the interstellar medium. Their unique spectra contain information about the quality and quantity of the ionizing field as well as the temperature, density, and metallicity of these discrete locations in the Galaxy. With increasing sensitivity across many spectral domains, we now know that ionized gas is not localized to massive star regions in many star-forming galaxies. In particular, recent observational studies allow a thorough comparison of the physical conditions and distribution of the well-studied classical H II regions to the more widespread warm, diffuse gas. By more realistically evolving a dynamic interstellar medium, models are beginning to reproduce the observed emission measure variations and provide a natural solution to the propagation of ionizing flux from a predominantly neutral galactic disk to the distant halo.
\end{abstract}

Bright, ionized nebulae are one of the most obvious sites where feedback from massive stars impacts the interstellar medium (ISM) in star-forming galaxies. But the discovery and study of widespread ionized gas over a wide range of temperatures ($T \sim 10^4$--$10^6$ K) throughout the last century reveals that energy is deposited through other, large-scale processes. One phase that has received considerable attention over the last four decades is the pervasive, warm ($\sim 0.7$--$1.2 \times 10^4$ K) ionized medium (WIM). In most spiral galaxies with star formation similar or more vigorous than the Milky Way, the WIM is a thick ($\sim$ few kpc) layer of gas with a high fraction ($\gtrsim 80\%$) of ionized hydrogen. \citet{HafDetBec09} reviews much of the observational and theoretical work done recently to define its properties and explore the possibilities for powering the ionization and heating of this component. Here, we review the history of early observations, issues that arose from those first data, and early simulations that explored solutions (\S\ref{sec:discovery} and \S\ref{sec:early_models}); more recent extragalactic and Galactic observations made possible with modern instrumentation (\S\ref{sec:extragalactic} and \S\ref{sec:surveys}); and new simulations inspired by a growing appreciation for turbulence and the complex density distribution of a real, dynamic ISM (\S\ref{sec:simulations}).

\section{Discovery \&  Early Exploration}
\label{sec:discovery}

The existence of substantial ionized gas outside of classical H~\textsc{ii} regions was first suggested by \citet{HoyEll63}. Very low frequency radio observations of the Galactic synchrotron emission show a turnover below 10 MHz, a marked deviation from a constant power law at higher frequencies. The peak emission in the spectrum changes with latitude, with the highest latitudes having the lowest frequency turnover. Hoyle \& Ellis interpreted this break from the power law relationship as evidence for free-free absorption of the synchrotron emission by electrons along the line of sight. With a clear lack of discrete ionized regions at high latitudes, they posited a diffuse ionized layer must exist to provide the absorption. Using the latitude dependence of the peak shift and plausible arguments about the temperature and density of the absorbing layer, they present a model of the WIM with physical characteristics only a factor of a few from adopted values today.

Shortly following their work, the discovery of pulsars \citep{HewBelPil68} afforded a direct measure of the free electron column along lines of sight since their dispersion measure (DM $\equiv \int n_{e}\,dl$) causes a time delay in pulse arrival as a function of frequency. Significant dispersion measures are seen toward pulsars---especially those above the plane---that have no obvious ionized region along the line of sight. 

While both of these observations from the radio sky provide strong evidence for ionized gas far from classical H~\textsc{ii} regions, they are difficult to fully examine the physical characteristics of the WIM. Instead, investigators pushed hard to detect direct emission from the WIM, most notably in the optical where the predicted densities and temperatures of the gas were expected to produce a very faint spectrum similar to H~\textsc{ii} regions, including the typical hydrogen recombination lines. \citet{Siv74} presented a deep, wide-field H$\alpha$ survey that revealed substantially fainter ionized regions than had been previously seen, including faint, extended structure near several nearby H~\textsc{ii} regions. However, detection of the WIM in moderate-resolution imaging surveys was not possible until the advent of modern CCD detectors (see \S\ref{sec:surveys}).

In the meantime, \citet{ReyRoeSch73,ReySchRoe73} utilized large-format (15 cm) Fabry-Perot etalons to detect and resolve spectral line emission from the ISM more than an order of magnitude fainter than the sensitivity limits of direct imaging at the time. Leveraging the ability of such spectrometers to accept a large solid angle, Reynolds traded spatial resolution ($\sim1\deg$) for sensitivity and spectral resolution and began an observational campaign to fully explore the nature of this extended, pervasive ionized component of the ISM.

After about two decades of dedicated work, Reynolds and collaborators had established the basic characteristics of the WIM in the Milky Way:

\begin{itemize}
\item Locally, in a vertical column through the solar neighborhood, \\
	$H^+ \approx {1\over3} H^0: N_{H^+} \approx 1 \times 10^{20} \mathrm{cm}^{-2}$.
\item Globally, 90\% of the H$^+$ mass is in the WIM, not in H II regions.
\item Fills 20--40\% of the volume over a 2--3 kpc thick layer about the midplane.
\item Requires $1 \times 10^{-4}$ erg s$^{-1}$ cm$^{-2}$ to sustain the $5 \times 10^6$ s$^{-1}$ cm$^{-2}$ recombination rate.

\end{itemize}
	
This power rate is quite high \citep{Rey84}. In comparison to some well-known sources, the observed recombination rate requires nearly all the mechanical energy output from supernovae in the Galaxy or about 15\% of the total ionizing flux of massive (OB) stars. While the latter provides a much more comfortable margin and a natural extension of their obvious H~\textsc{ii} regions, there are two immediate problems that need to be explained.

First, if massive stars are the source, we must be able to explain how at least 15\% of their ionizing flux escapes to these large distances above the plane. With their short lifetimes and the concentration of massive star-forming regions close to the plane, few of these stars travel far from the dense medium of their birth. Combined with the large cross section of neutral hydrogen, simple distributions of the global ISM do not allow this population to contribute much at large distances. Some early attempts to add realistic complexity to models are discussed in \S\ref{sec:early_models}, while more recent efforts leveraging our tremendous increase in resources today are reviewed in \S\ref{sec:simulations}.

Second, while the WIM does emit optical forbidden line emission similar to O-star H~\textsc{ii} regions, it has characteristically \emph{lower} [O~\textsc{iii}]/H$\alpha$ and He~\textsc{i}/H$\alpha$ ratios as well as \emph{higher} [N~\textsc{ii}]/H$\alpha$ and [S~\textsc{ii}]/H$\alpha$. The simplest interpretation of these differences is that lower ionization states are maintained in the WIM; the dominant ions are N$^+$, S$^+$, and O$^+$. Such a result could be explained by a softer ionizing spectrum, however as ionizing flux is absorbed by neutral hydrogen away from the source it typically \emph{hardens}; photons near 13.6 eV are preferentially absorbed. More likely, as shown by \citet{DomMat94}, the gas can equilibrate in these lower states if the photon to gas density ratio (the ionization parameter) is very low. In such a dilute radiation field, atoms have time to recombine on average before the next ionizing photon arrives. 

\section{Early Models}
\label{sec:early_models}

While more comfortable solutions exist to these two problems today (as described later), the observational findings spawned a number of creative solutions to provide a more extended or even \emph{in situ} ionization source. Some ideas included:

\begin{itemize}
\item Stellar components with larger scale heights: B-stars, evolved cores, \& white dwarfs \citep{Rey84}.
\item Supernovae through cooling hot gas, shocks, mixing layers \citep{SlaMcKHol00,SlaShuBeg93,ShaBen91}.
\item Magnetic reconnection powered by supernovae or galactic rotation \citep{Ray92}.
\item Decaying neutrinos \citep{Sci90}.
\end{itemize}

Although the power requirements still make it hard for one of these options to dominate the ionization of the WIM, many of these sources (save the last) are likely to contribute at some level in spiral galaxies. Some of the variation in line ratios, especially at large distances from the plane, may be due to a higher percentage of flux from these sources in the halo \citep[e.g.,][]{ColRan01}. 

At the same time that alternative ionizing sources were being investigated, several groups turned their attention to examining the details of Lyman continuum (LyC) transport. \citet{MilCox93} describe a scenario where an originally smooth medium populated with opaque clouds is ionized by massive stars to create the WIM from large-scale, overlapping H II regions. Focusing instead on concentrations of massive stars to break out of the disk, \citet{DovShu94} show that OB associations are density bounded vertically, allowing radiation to escape to large distances. \citet*{BasJohMar99} model a specific dynamical solution for the superbubble associated with the W4 H~\textsc{ii} region, which \citet*{DenTopSim97} show has associated faint H$\alpha$. The specific model that explains the dynamical size of the superbubble and the associated emission also allows about 15\% of the original ionizing flux to escape vertically from the region. \citet*{DovShuFer00} further explore this idea more generally for associations, examining larger vertical regions and exploring various time evolution scenarios for the star formation history.

The collection of these transport studies demonstrated that ionizing flux can be propagated to large distances above the plane with reasonable conditions: either within an association and/or with a specific distribution of the neutral medium. Whether these conditions are justified and globally applicable to explain the pervasive nature of the WIM has only recently been able to be studied in new simulations of the fully multiphase and dynamic ISM (see \S\ref{sec:simulations}).

\section{An Extragalactic Perspective}
\label{sec:extragalactic}

As with many topics dependent on optical astronomy, the introduction of CCDs into mainstream research in the early 1990s propelled the study of the WIM to a new level. The increased sensitivity and linearity allowed new investigations of the morphology and extent of faintly emitting gas in the Milky Way (\S\ref{sec:surveys}), providing a new view of the relationship between H~\textsc{ii} regions and the diffuse gas. The WIM in other spiral galaxies could also be studied to similar physical extents from the plane as our galaxy, showing that this component of our ISM was ubiquitous in similar galaxies. 

\begin{figure}[t]
\begin{center}

\plotone{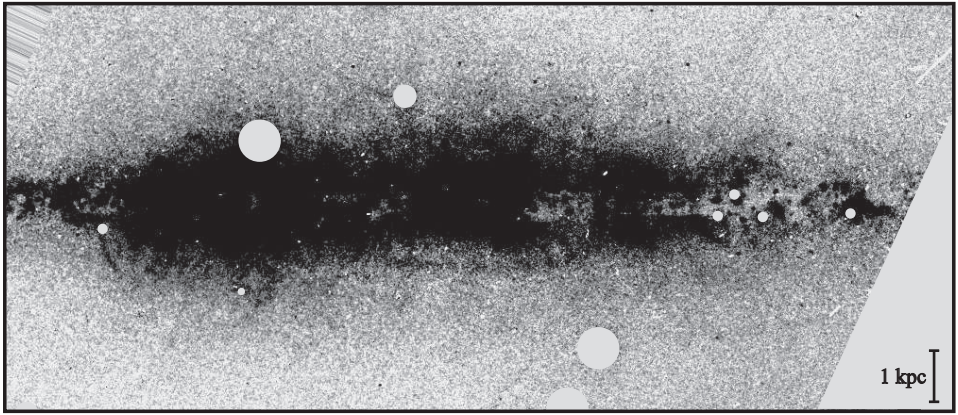}
\caption{\textbf{The WIM in NGC 891.} A continuum-subtracted H$\alpha$ image of this nearly edge-on galaxy. Adapted from \citet{HowSav00}.}
\label{fig:ngc891}

\end{center}
\end{figure}

Deep imaging and spectroscopy of the diffuse gas of many spiral and irregular star-forming galaxies provided a welcome global view in the early 1990s. Since the fainter emitting gas in other galaxies was not necessarily known \emph{a priori} to be similar to the WIM of the Milky Way, extragalactic researchers called it ``diffuse ionized gas'' (DIG). Today it is generally recognized that most of the DIG emission traces the WIM component in a galaxy. H$\alpha$ imaging of edge-on spirals reveals a large, extended, diffuse ionized component that generally matches the inferred distribution of the Milky Way from our interior vantage point (Figure~\ref{fig:ngc891}). Complementary observations of face-on galaxies show diffuse emission following star formation and the general structure of the spiral arms. Spectroscopy of the diffuse gas in a variety of galaxy types and orientations replicate general trends in the Milky Way but exhibit variation in metallicity and excitation present in such a larger observational sample.

The deluge of observations from these initial studies provided a global view that further reinforced the WIM as a distinct component of the ISM in spirals:

\begin{itemize}
\item 25--60\% of H$\alpha$ luminosity emitted from spiral galaxies comes from their WIM, not from their H~\textsc{ii} regions. The power requirement is quite large and on scale with that required for the Milky Way.
\item As in the Milky Way, the WIM in spirals is generally distributed as a wide-spread, thick disk of H$^+$ with scale heights typically 1--3 kpc. However this new perspective revealed some radial variation in scale height in some galaxies \citep[e.g.,][]{Ran96,HooWalRan99}.
\item Metal ions are also preferentially in lower ionization states, clearly now a unique spectral tracer of the WIM. Line ratios of these dominant ions to H$\alpha$ vary inversely with H$\alpha$ intensity, as in our galaxy \citep[e.g.,][]{OttGalRey02,OttReyGal01,ColRan01,TulDet00,GolDetDom96}.

\end{itemize}

\citet{RosDet03a,RosDet03} completed one of the most comprehensive H$\alpha$ imaging surveys of spirals to date. Their study shows a clear link between infrared tracers of active star formation and the existence of an extended ionized layer in a galaxy. Such a relationship suggests that either the source responsible for powering the star formation or a byproduct of the formation itself is responsible for maintaining the WIM layer. Although this link is consistent with ionization by massive stars, it does not rule out a role played by other sources such as supernovae. 

Evidence for a closer link between the massive stars and the WIM has come from the work of \citet{FerWysGal96} and \citet*{ZurRozBec00}. In these studies of the ionized gas in moderately inclined to face-on spirals they trace lower-limit WIM-to-total H$\alpha$ luminosity fractions of 25--50\% over the whole disk. Furthermore, they show that these fractions are roughly constant as a function of radius for the galaxies in their samples despite the decreasing star-formation rate per unit area. In their very sensitive observations, \citet{ZurRozBec00} observe that enhanced diffuse emission near H~\textsc{ii} regions supports that LyC flux is leaking from these sites. This impression persists despite the fact that the extremely high contrast and sharp boundary between an H~\textsc{ii} region and the ``background'' suggests that the ionized region should be photon bounded (i.e., the flux has been completely absorbed). 

\citet{ZurBecRoz02} take this notion one step further and attempt to model the DIG of NGC 157 beginning with the actual H~\textsc{ii} region distribution and luminosities as well as the actual H~\textsc{i} distribution based on 21 cm observations. They are able to examine a variety of escape fraction scenarios and compare the results to their deep H$\alpha$ observations. While the results do not quite match the resolution of their optical observations due to limitations in the 21 cm data, the general agreement lends additional support to a ``leaky'' H~\textsc{ii} region hypothesis. 

\section{Galactic H\boldmath$\alpha$ Imaging Surveys and WHAM}
\label{sec:surveys}

\begin{table}[t]
\caption{Recent H$\alpha$ Surveys}
\label{table:surveys}
\smallskip
\begin{center}
{\small
\begin{tabular}{lcccc}
\tableline
\noalign{\smallskip}
Survey & Coverage & Sensitivity & Resolution & Notes\\
\noalign{\smallskip}
\tableline
\noalign{\smallskip}
Wisconsin H-Alpha& complete & $< 0.1$ R & $1\deg$ & (1) \\
Mapper (WHAM) & & & \\
\noalign{\medskip}
Southern H-Alpha Sky & $\delta < +15\deg$ & few R & few arcmin & (2) \\
Survey Atlas (SHASSA) & & \\
\noalign{\medskip}
Virginia Tech Spectral-line & $\delta > -15\deg$ & few R & few arcmin & (3) \\
Survey (VTSS) & $|b| < 30\deg$ & \\
\noalign{\medskip}
AAO/UKST SuperCOSMOS & $\delta > +2\deg$ & $< 5$ R & 1--2 arcsec & (4) \\
H-Alpha Survey (SHS) & $|b| < 10\deg$ & \\
\noalign{\medskip}
INT/WFC Photometric  & $\delta > +10\deg$ & few R & $< 2$ arcsec & (5) \\
H-Alpha Survey (IPHAS) & $|b| < 5\deg$ & \\
\noalign{\medskip}
VST/OMEGACAM Photometric & $\delta < -10\deg$ & few R & $< 2$ arcsec & (6) \\
H-Alpha Survey (VPHAS+) & $|b| < 5\deg$ & \\
\noalign{\smallskip}
\tableline
\end{tabular}
}
\end{center}
{\footnotesize
(1) \citealt{WHAMNSS}; \url{http://www.astro.wisc.edu/wham/}. Survey is spectral, not imaging. For details and southern survey progress, see \S\ref{sec:surveys}. All-sky public release expected in 2011 or 2012.\\
(2) \citealt{SHASSA}; \url{http://amundsen.swarthmore.edu/}.\\
(3) \citealt{VTSS}; \url{http://www.phys.vt.edu/~halpha/}. About 50\% of fields available.\\
(4) \citealt{UKST-SC}; \url{http://www-wfau.roe.ac.uk/sss/halpha/}. Digitally scanned from originally photographic survey.\\
(5) \citealt{IPHAS}; \url{http://www.iphas.org/}. About 60\% available in an initial data release; fully calibrated survey to be released near end of 2010.\\
(6) \url{http://www.vphas.org/}. Preliminarily approved for the VLT Survey Telescope with observations expected to start in 2011.\\
}
\end{table}

\begin{figure}[t]
\begin{center}

\plotone{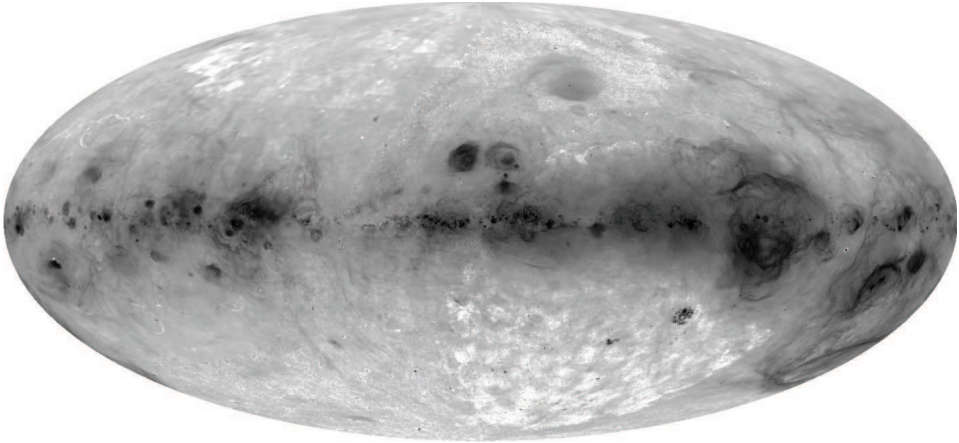}
\caption{\textbf{A composite view of the H$\alpha$ sky.} WHAM, SHASSA, and VTSS surveys combine to provide a sensitive, all-sky picture of the diffuse ionized gas of the Milky Way. Data from \citet{Fin03}.}
\label{fig:ha-sky}

\end{center}
\end{figure}

The advance in detector technology also spurred several groups to initiate deep, comprehensive surveys of H$\alpha$ from the Milky Way. Table~\ref{table:surveys} compiles a list of these recent efforts with their characteristics. WHAM and SHASSA are the only all-sky surveys to date; the other four surveys listed concentrate on the Galactic plane. \citet{Fin03} has combined WHAM, SHASSA, and VTSS into a complete Galactic survey (Figure~\ref{fig:ha-sky}), although the resolution and sensitivity vary across the image. While most of these new surveys are imaging, our own efforts have been focused on providing a unique spectroscopic view.

\begin{figure}[t]
\begin{center}

\plotone{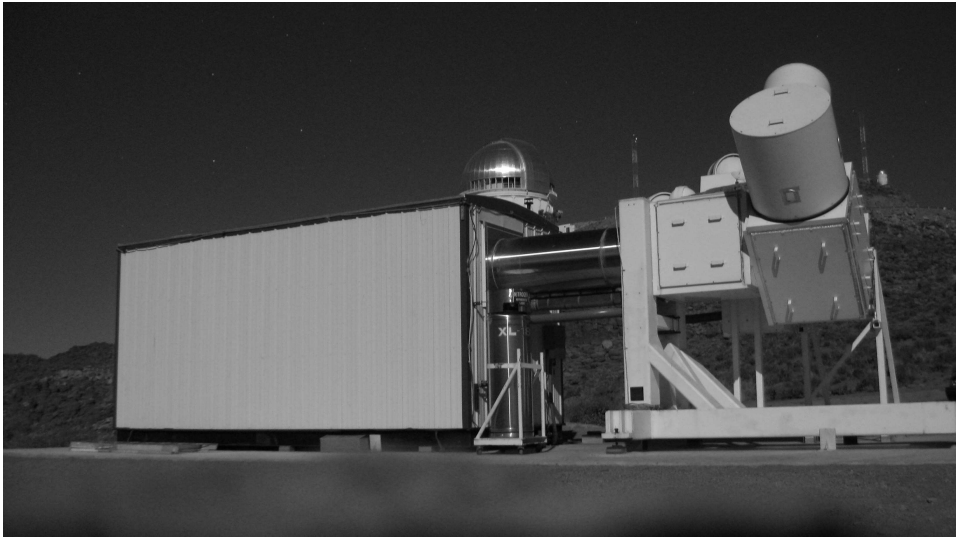}
\caption{\textbf{The Wisconsin H-Alpha Mapper.} WHAM observing the Chilean sky from Cerro Tololo in early 2009. Photo courtesy Alex Hill.}
\label{fig:WHAM}

\end{center}
\end{figure}

Leveraging the success of the Fabry-Perot spectrometer for study of the WIM, in the early 1990s Reynolds proposed to build a dedicated observatory to undertake the first kinematic survey of the diffuse ionized gas in the Galaxy. With funding primarily from the National Science Foundation, he and collaborators built the Wisconsin H-Alpha Mapper (WHAM) to be highly optimized for observations of large-angular-scale optical emission (Figure~\ref{fig:WHAM}). WHAM consists of an all-sky siderostat that feeds a 0.6-m primary lens that delivers a 1\deg\ beam on the sky to a dual-etalon, 15-cm diameter Fabry-Perot spectrometer. The primary configuration produces a 200 km~s$^{-1}$ spectrum with 12 km~s$^{-1}$ resolution. Gap spacings are fixed in the etalons, but this spectral window can be tuned between about 4800 \AA\ and 7300 \AA\ by changing the gas pressure (SF$_6$) in the etalon chambers. Considerable detail about WHAM's optical design and performance can be found in \citet{TuftePhD}.

Commissioned at Pine Bluff Observatory (PBO) in Wisconsin during 1996, we installed WHAM at Kitt Peak National Observatory (KPNO) in November, 1996 and began the Northern Sky Survey \citep[WHAM-NSS; ][]{WHAMNSS} in early 1997. Two years of H$\alpha$ survey observations were followed by an exploration of the physics of the WIM through other optical emission lines \citep[e.g.,][]{MadReyHaf06,MadRey05,HauReyHaf02,ReySteHaf01a,ReyHafTuf99,HafReyTuf99,ReyHauTuf98} and attempts to detect and map faint H$\alpha$ emission from high-velocity complexes \citep[e.g.,][]{HafDunHof09,Haf05,TufWilMad02,HafReyTuf01b,TufReyHaf98}. After eleven years at KPNO, we decommissioned WHAM and moved it back to PBO for a year of maintenance. In early 2009, WHAM shipped to Chile and was installed at Cerro Tololo Inter-American Observatory (CTIO) in March. After a short period of recalibration and recommissioning, science observations resumed. At this point (mid-2010), more than 90\% of the H$\alpha$ data newly observable ($\delta < -30\deg$) and more than 70\% of a full southern sky survey ($\delta < +30\deg$) has been obtained. We expect to release a fully calibrated, all-sky survey by late 2011 or early 2012. 

While surveys of neutral hydrogen have typically enjoyed high spectral resolution \citep[e.g.,][]{GASS,LAB,IAR,LDS}, WHAM delivers the first all-sky kinematic survey at H$\alpha$. Although the angular resolution is modest for optical wavelengths, its sensitivity and ability to separate terrestrial lines from Galactic components reveals emission that covers the sky. Spectral resolution also allows us to isolate structure along lines of sight where rotation separates spiral arms and provides new insight into the dynamics of large-scale discrete objects. An example of the power of the kinematic survey can be found in Haffner et al.\ (this volume, pg.\ xxx).


\section{New Simulations}
\label{sec:simulations}

Simulations examining the structure and powering of the WIM and its connection to H~\textsc{ii} regions have also progressed significantly over the past few years. Beyond the continuing increase in computational resources, a resounding theme in recent years has been the impact of density variations on ionized regions. Recognizing that dynamics in either organized or turbulent forms has a considerable impact on the density structure of the ISM, many recent works are finding that such conditions change the ionization and emission of regions to better match recent observations. 

Examining the internal structure of locally ionized regions, \citet{GiaBecZur04, GiaBecCed05} and \citet{WooHafRey05} find that inhomogeneities in H~\textsc{ii} regions impact not only the internal ionization and emission structure, but also the amount of ionizing flux that escapes. In the future, these results can be linked more directly to global simulations of the WIM that include input from actual H~\textsc{ii} region distributions, such as that of \citet{ZurBecRoz02}, discussed in \S\ref{sec:extragalactic} above. 

At larger scales, \citet{HilBenKow08} find that the emission distribution of the WIM is lognormal, which leads them to compare the MHD models of a turbulent ISM produced by \citet{KowLazBer07}. A simple, idealized simulation cube allows them to explore a wide range of parameter space in this study. Simulations that best match the observed emission distribution are mildly supersonic ($\mathcal{M} \sim 1.4$--2.4), producing line widths consistent with those observed by WHAM. They also find that the simulated emission distribution is relatively insensitive to changes in the magnetic field strength. 

Using more complex conditions that emulate a slice of the Galaxy, \citet{WooHilJou10} ionize the supernovae-driven turbulent medium generated in the models of \citet{JouMac06} and \citet*{JouMacBry09}. Although the additional complexity does not permit them to study a large parameter space, the attempt to match galactic conditions provides an interesting complement to the more idealized models. They show that with an ionizing source distribution similar to that near the sun, LyC can travel to large distances ($> 2$ kpc) from the plane. While the details of the gas distributions do not match observations in these non-magnetic simulations, the ability to ionize the WIM through a self-consistent, dynamic ISM is encouraging. Hill et al.\ (this volume, pg.\ xxx) discuss this latest work in more detail.


\section{Summary}

Recent observations and simulations suggest that a dynamic ISM is essential to explain the ionization and emission structure of the WIM. A turbulent, fractal medium allows LyC radiation to escape H~\textsc{ii} regions and ionize gas far from massive stars. These elements may soon allow a fully global model to solidify the link between active star formation and powering the WIM.

WHAM will soon finish the southern component of the first all-sky kinematic H$\alpha$ survey of the Milky Way. Followup with other emission line surveys will allow us to compare physical conditions of the WIM in a variety of conditions across the whole Galaxy. WHAM will then turn its gaze farther toward the Magellanic system to explore diffuse ionization in the unique conditions provided, in particular, by the extended structures of the Bridge and Stream.

\acknowledgements LMH and WHAM are supported by NSF award AST-0607512. WHAM was built with the help of the University of Wisconsin Graduate School, Physical Sciences Lab, and Space Astronomy Lab. Their staff and that of KPNO and CTIO have greatly contributed to the success of the project. 

\bibliography{haffner1_l}

\end{document}